\documentclass[aps,prl,superscriptaddress,amsfonts,amsmath,amssymb,showpacs,floatfix,reprint]{revtex4-1}

\usepackage{bm}
\usepackage{graphicx}
\usepackage{amsmath}
\usepackage{amstext}
\usepackage{amssymb}
\usepackage{amsfonts}
\usepackage{amsbsy}
\usepackage{verbatim}
\usepackage[usenames,dvipsnames,svgnames,table]{xcolor}
\usepackage{multirow}
\usepackage{gensymb}
\usepackage{textcomp}
\usepackage{enumitem}
\usepackage{comment}
\usepackage{fancyhdr}
\usepackage[version=3]{mhchem}
\usepackage[colorlinks=true, urlcolor=Blue, linkcolor=Blue, citecolor=Blue, pdftex]{hyperref}
\usepackage{times}
\usepackage{nicefrac}
\usepackage{dsfont}

\newcommand{\be}{\begin{equation}}
\newcommand{\ee}{\end{equation}}

\newcommand{\bee}{\begin{equation*}}
\newcommand{\eee}{\end{equation*}}

\newcommand{\1}{\hspace*{-1pt}}
\newcommand{\2}{\hspace*{-2pt}}
\newcommand{\3}{\hspace*{-3pt}}

\begin{document}
\title{Characterization of quantum spin liquids and their spinon band structures via functional renormalization}
\author{Max Hering}
\affiliation{Dahlem Center for Complex Quantum Systems and Institut f\"ur Theoretische Physik, Freie Universit\"{a}t Berlin, Arnimallee 14, 14195 Berlin, Germany}
\affiliation{Helmholtz-Zentrum f\"{u}r Materialien und Energie, Hahn-Meitner-Platz 1, 14019 Berlin, Germany} 
\author{Jonas Sonnenschein}
\affiliation{Dahlem Center for Complex Quantum Systems and Institut f\"ur Theoretische Physik, Freie Universit\"{a}t Berlin, Arnimallee 14, 14195 Berlin, Germany}
\affiliation{Helmholtz-Zentrum f\"{u}r Materialien und Energie, Hahn-Meitner-Platz 1, 14019 Berlin, Germany} 
\author{Yasir Iqbal}
\affiliation{Department of Physics, Indian Institute of Technology Madras, Chennai 600036, India}
\author{Johannes Reuther}
\affiliation{Dahlem Center for Complex Quantum Systems and Institut f\"ur Theoretische Physik, Freie Universit\"{a}t Berlin, Arnimallee 14, 14195 Berlin, Germany}
\affiliation{Helmholtz-Zentrum f\"{u}r Materialien und Energie, Hahn-Meitner-Platz 1, 14019 Berlin, Germany} 

\date{\today}

\begin{abstract}
We combine the pseudofermion functional renormalization group (PFFRG) method with a self-consistent Fock-like mean-field scheme to calculate low-energy effective theories for emergent spinon excitations in spin-1/2 quantum spin liquids. Using effective spin interactions from PFFRG as an input for the Fock equation and allowing for the most general types of free spinon ans\"atze as classified by the projective symmetry group (PSG) method, we are able to systematically determine spinon band structures for spin-liquid candidate systems beyond mean-field theory. We apply this approach to the antiferromagnetic $J_1$-$J_2$ Heisenberg model on the square lattice and to the antiferromagnetic nearest-neighbor Heisenberg model on the kagome lattice. For the $J_1$-$J_2$ model, we find that in the regime of maximal frustration a SU(2) $\pi$-flux state with Dirac spinons yields the largest mean-field amplitudes. For the kagome model, we identify a gapless $\mathds{Z}_2$ spin liquid with a small circular spinon Fermi surface and approximate Dirac-cones at low but finite energies.
\end{abstract}
\maketitle

{\em Introduction}.
The investigation of fractional quasiparticles has developed into one of the most active research topics in modern condensed-matter physics. In general, fractionalization occurs whenever the excitations of a many-body system carry quantum numbers which are fractions of those of the actual elementary constituents. Historically, such \emph{emergent} phenomena were first theoretically discussed for certain one-dimensional interacting electron systems showing spin-charge separation~\cite{Luther-1974,Su-1979,*Su-1981,Haldane-1981,Kim-2006}. Later, the observation of the fractional quantum Hall effect in two-dimensional electron gases marked the experimental breakthrough in this growing research area~\cite{Tsui-1982,Laughlin-1983,Halperin-1984}. Lately, fractionalization has attracted increasing interest in the context of magnetic systems where, among other examples~\cite{,Sen-2011,Rehn-2017}, it manifests through monopole excitations in classical spin ice systems~\cite{Castelnovo-2008,Morris-2009,Fennell-2009,Kadowaki-2009} or through spinons in spin-$1/2$ quantum spin liquids~\cite{Mourigal-2013,Han-2012}.

In the latter scenario, strong magnetic frustration effects and large quantum fluctuations hinder a spin system from developing conventional long-range magnetic order~\cite{Anderson-1973}. As a consequence, the bosonic $S=1$ spin-wave excitations of an ordered state can decompose into fractional and deconfined quasiparticles with spin $S=1/2$, called spinons~\cite{Faddeev-1981,Hao-2009}. The emergent nature of this effect is directly evident when assuming a model with local spin operators $\mathbf{\hat{S}}_i$ such as a Heisenberg Hamiltonian
\begin{equation}
\mathcal{\hat{H}}=\sum_{(i,j)}J_{ij}\mathbf{\hat{S}}_{i}\cdot\mathbf{\hat{S}}_{j}\label{heisenberg}
\end{equation}
with pairs of lattice sites labelled $(i,j)$. Even though the spin operators are only capable of changing local spin quantum numbers by integer multiples of Planck's constant ($\Delta S=0,\pm1$), a quantum spin liquid ground state nevertheless features excitations with spin-$1/2$~\cite{Faddeev-1981,Hao-2009}. In two-dimensions, the existence of fractionalized excitations in quantum spin liquids is well established for only a handful of systems, namely, the $S=1/2$ Kalmeyer-Laughlin abelian-~\cite{Kalmeyer-1987,*Kalmeyer-1989} and a $S=1$ nonabelian chiral spin liquid~\cite{Greiter-2009,*Greiter-2014}, the resonating-valence bond phase in the quantum dimer model on the triangular lattice~\cite{Moessner-2001}, and the Kitaev model spin liquid~\cite{Kitaev-2006}. Beyond one-dimension, the low-energy theory of quantum spin liquids also requires a coupling of the spinons to an emergent gauge field which may enrich the system with anyonic quasiparticle statistics~\cite{Wilczek-book} and topologically protected ground state degeneracies~\cite{Wen-1990}.

The theoretical prediction of fractionalization in quantum many-body systems starting from a `bare' Hamiltonian is a notoriously difficult problem. Particularly, for a generic frustrated Heisenberg model, already the numerical identification of a spin-liquid ground state poses a significant challenge. The characterization of its emergent excitations (such as spinon band structures and the type of gauge field they are associated with) represents an even more difficult endeavor. Important but somewhat indirect insight into spinon properties can be gained from the scaling of the entanglement entropy calculated by DMRG~\cite{Depenbrock-2012,Jiang-2012b} (which recently even allowed for a qualitative investigation of emergent Dirac spinons in the kagome spin liquid~\cite{zhu18}). Furthermore, variational Monte Carlo (VMC) allows to identify the free fermion model with the lowest variational energy of its Gutzwiller-projected ground state, however, this free fermion model does not necessarily describe the system's spinon excitations (yet, it has recently met with some success in describing spinon excitations~\cite{Ferrari-2018a,Ferrari-2018b}). 

In this letter we develop and apply a numerical technique based on PFFRG~\cite{reuther10} which {\it directly} calculates low-energy effective theories of fermionic spinons in a quantum spin liquid. Within our method, we first calculate renormalized spin interactions via one-loop PFFRG by integrating out energy modes of Eq.~(\ref{heisenberg}) down to a low but finite scale $\Lambda$. These effective interactions are then used in a Fock-like mean-field scheme treating the remaining low-energy modes and allowing us to determine the kinetic terms of an effective spinon theory (including spinon hopping and \emph{singlet} pairing). This approach combines the strengths of PFFRG which captures significant parts of the system's quantum fluctuations with the advantages of a mean-field treatment wherein emergent spinon properties (that would not follow from a PFFRG calculation alone) may be determined self consistently (see also Refs.~\cite{reiss07,wang14} for related works on Hubbard models). We apply our approach to spin-liquid candidate systems on the square and kagome lattices where the possible {\it ans\"atze} for spinon hopping and pairing are taken from the respective PSG classifications~\cite{wen02,Lu-2011}. Our findings are discussed and benchmarked against VMC results.

{\em Method}.
Our starting point is the fermionic parton construction of spin operators~\cite{Abrikosov-1965} ${\mathbf {\hat{S}}}_i=\frac{1}{2}\hat{f}_{i}^{\dagger}\bm{\sigma}\hat{f}_{i}$ with $\hat{f}_{i}=(\hat{f}_{i\uparrow},\hat{f}_{i\downarrow})$ which, in a first direct approach, may be inserted into Eq.~(\ref{heisenberg}) to yield a purely quartic Hamiltonian. Such types of `bare' fermionic Hamiltonians have recently been successfully treated by PFFRG, allowing for an unbiased investigation of magnetically ordered and disordered phases in 2D~\cite{Reuther-2011a,Reuther-2011b,Reuther-2011c,Singh-2012,Reuther-2014,Suttner-2014,Iqbal-2015,Iqbal-2016a,Iqbal-2016c,Buessen-2016,Hering-2017,Baez-2017,Buessen-2018b,Roscher-2018,Keles-2018} and 3D spin systems~\cite{Iqbal-2016b,Balz-2016,Buessen-2016,Iqbal-2017,Buessen-2018a,Rueck-2018,Chillal-2017,Iqbal-2018a,Iqbal-2018b}. Within this technique, the free fermion propagator $G_0(\omega)=\frac{1}{i\omega}$ is regularized by a step-like infrared cutoff function for Matsubara frequencies, yielding $G_0^\Lambda(\omega)=\frac{\Theta(|\omega|-\Lambda)}{i\omega}$. The artificial $\Lambda$ dependence of this modified theory can be described by a hierarchy of coupled integro-differential equations for all $m$-particle vertex functions $\Gamma_m^\Lambda$. Solving these equations using the standard one-loop plus Katanin truncation yields renormalized and cutoff-dependent vertices $\Gamma_1^\Lambda$ and $\Gamma_2^\Lambda$. Here, $\Gamma_1^\Lambda\equiv\Sigma^\Lambda$ is an onsite imaginary self energy (i.e., an inverse fermionic lifetime) and $\Gamma_2^\Lambda\equiv\Gamma^\Lambda$ is a frequency-dependent effective spin interaction which typically becomes more spread out in real space compared to the range of the bare interactions $J_{ij}$ recovered at $\Lambda\rightarrow\infty$. It is important to emphasize that in the absence of quadratic terms in the fermionic Hamiltonian, the self energy $\Sigma^\Lambda$ remains strictly local during the RG flow.

Alternatively, the parton construction may be used to formulate an effective low-energy theory for spinons which (for $\mathds{Z}_2$ quantum spin liquids) reads
\begin{equation}
\mathcal{\hat{H}}_{\text{eff}}=\sum_{(i,j)}(\hat{\psi}_i^\dagger u_{ij}\sigma_{ij}^z\hat{\psi}_j+\text{h.c.})\,.\label{effective}
\end{equation}
Here $\sigma_{ij}^z=\pm1$ are gauge degrees of freedom living on the lattice bonds [in the case of U(1) or SU(2) spin liquids these fields have the form of a complex phase or are of SU(2)-matrix-type, respectively] and $\hat{\psi}_{i}=(\hat{f}_{i\uparrow},\hat{f}_{i\downarrow}^\dagger)^{T}$. Expanding $u_{ij}$ in terms of Pauli matrices $\tau^\mu$ (where $\mu=0,1,2,3$ and $\tau^0$ is the identity matrix) the coefficients of $\tau^3$ ($\tau^1$) correspond to real spinon hopping $\text{Re}[t_{ij}]$ (real singlet pairing $\text{Re}[\Delta_{ij}]$). Also note that the partons establish a local gauge freedom~\cite{Baskaran-1988} according to which the transformation $\hat{\psi}_{i}\rightarrow W_{i}\hat{\psi}_{i}$ leaves the physical spin operators $\mathbf{\hat{S}}_i$ unchanged [$W_i$ is a $2\times2$ SU(2) matrix]~\cite{Affleck-1988,Dagotto-1988}.

The key difficulty associated with this effective theory is that it is {\it a priori} unclear how exactly Eq.~(\ref{effective}) arises from the aforementioned bare fermionic Hamiltonian. According to the more traditional approach~\cite{Baskaran-1987} of obtaining Eq.~(\ref{effective}), the terms $(\hat{f}_i^\dagger\bm{\sigma}\hat{f}_i)(\hat{f}_j^\dagger\bm{\sigma}\hat{f}_j)$ may be mean-field decoupled into $t_{ij}\hat{f}_{i\alpha} \hat{f}_{j\alpha}^\dagger$ with $t_{ij}\sim J_{ij}\langle \hat{f}_{i\alpha}^\dagger \hat{f}_{j\alpha}\rangle$ and into $\Delta_{ij}\hat{f}_{i\uparrow}^\dagger \hat{f}_{j\downarrow}^\dagger$ with $\Delta_{ij}\sim J_{ij}\langle \hat{f}_{i\downarrow} \hat{f}_{j\uparrow}\rangle$. In Feynman many-body language, this decomposition can be easily formulated for hopping and pairing simultaneously, yielding
\begin{equation}
u_{ij}=-\frac{3J_{ij}}{4\beta}\sum_{\omega}G_{ij}(\omega)e^{i\omega\delta},\text{ with }u_{ij}=\left(\begin{array}{cc}t_{ij}^*&\Delta_{ij}\\ \Delta_{ij}^*&-t_{ij}\end{array}\right)\,,\label{bare_mf}
\end{equation}
where $\omega$ denotes Matsubara frequencies and the propagator $G_{ij}(\omega)$ is a $2\times2$ matrix with anomalous contributions in the off-diagonal elements.
The self consistency for the amplitudes $u_{ij}$, which act as an effective Fock self energy, is closed via Dyson's equation $G_{ij}(\omega)=[G_0(\omega)^{-1}-u]^{-1}_{ij}$.

While the bare mean-field treatment in Eq.~(\ref{bare_mf}) provides first important insight into emergent spinon properties (see, e.g., the influential works~\cite{Baskaran-1987,affleck88, marston89,wen91,ubbens92,wen02}), obviously, its applicability is limited as it misses the gauge fluctuations described by $\sigma_{ij}^z$. Here, we substantially extend and generalize Eq.~(\ref{bare_mf}) by incorporating quantum fluctuations contained in the PFFRG vertices $\Sigma^\Lambda$ and $\Gamma^\Lambda$. Firstly, to adjust the regularization procedure of Eq.~(\ref{bare_mf}) to the one used in PFFRG we set the temperature $T$ to zero (i.e. $\frac{1}{\beta}\sum_{\omega}\rightarrow\frac{1}{2\pi}\int d\omega$) and instead use the regulator $\Lambda$ by replacing $G_0\rightarrow G_0^\Lambda$. The combination of PFFRG and Fock-mean-field analysis is then achieved by first normally integrating the RG equations down to a small but finite scale $\Lambda$. At this cutoff (which is naturally given by the $\Lambda$ value where the self-consistent amplitudes $u_{ij}$ first become finite), we start solving the Fock equation for the remaining low-energy modes where we add $\Sigma^\Lambda$ in Dyson's equations and replace $J_{ij}$ by the renormalized, frequency-dependent and more real-space spread out vertex $\Gamma^\Lambda$ \footnote{In order to also include the effects of quantum fluctuations at small energy scales we keep $\Sigma^\Lambda$ and $\Gamma^\Lambda$ evolving when we solve the Fock equation down to $\Lambda\rightarrow0$}. This allows us to self-consistently determine the amplitudes $u_{ij}$ which do not follow from a PFFRG analysis alone. The full Fock equation in momentum space with the PFFRG vertices inserted reads
\begin{equation}
u_\mathbf{k}^\Lambda\2=\2-\frac{1}{2\pi}\2\int\3 d\omega\2\sum_\mathbf{q}\Gamma_{\mathbf{k}-\mathbf{q}}^\Lambda(\omega,\omega,0)[G_0^\Lambda(\omega)^{-1}-\Sigma^\Lambda(\omega)-u_\mathbf{q}^\Lambda]^{-1}.\label{full_mf}
\end{equation}
Here, we have parametrized the frequency dependence of the two-particle vertex $\Gamma^\Lambda_\mathbf{k}(s,t,u)$ by the usual transfer frequencies $s$, $t$, and $u$ as defined in Ref.~\cite{reuther10}.

The absolute values $|t_{ij}|$ and $|\Delta_{ij}|$ obtained from solving Eq.~(\ref{full_mf}) generally depend on the assumed {\it phase pattern} of hoppings and pairings on different lattice bonds. Under the assumption that the spin-liquid ground state respects all lattice symmetries (unless stated otherwise we also preserve time-reversal invariance) these patterns follow from a PSG analysis which classifies all possible symmetry-allowed and gauge-inequivalent ans\"atze $u_{ij}$. We solve Eq.~(\ref{full_mf}) individually for all $\mathds{Z}_2$ PSG ans\"atze and identify the one which yields the largest amplitudes $|t_{ij}|$ and $|\Delta_{ij}|$. Note that even though these PSGs are derived from a $\mathds{Z}_2$ gauge structure, we may still detect U(1) or SU(2) spin-liquid solutions if, e.g., only finite hopping but no pairing amplitudes (or \emph{vice versa}) are generated. If $t_{ij}$ and $\Delta_{ij}$ are both finite on a certain bond type, we use $\xi_{ij}=(|t_{ij}|^2+|\Delta_{ij}|^2)^{1/2}$ as a measure for their combined strength, since this is the invariant quantity when performing gauge-transformations on the bond $(i,j)$. Finally, we stress that the self-consistent equation -- which may be symbolically written as $u_{ij}=f_{ij}(\{u_{lm}\})$ -- is form-invariant under gauge transformations $u_{ab}\rightarrow W_a^\dagger u_{ab} W_b$ such that solutions may be obtained in {\it any} gauge convention. This follows from the gauge-transformed self-consistent equation $W_i^\dagger u_{ij}W_j=f_{ij}(\{W_l^\dagger u_{lm}W_m\})$ due to the property $f_{ij}(\{W_l^\dagger u_{lm}W_m\})=W_i^\dagger f_{ij}(\{u_{lm}\})W_j$.
\begin{figure}[t]
\includegraphics[width=0.83\linewidth]{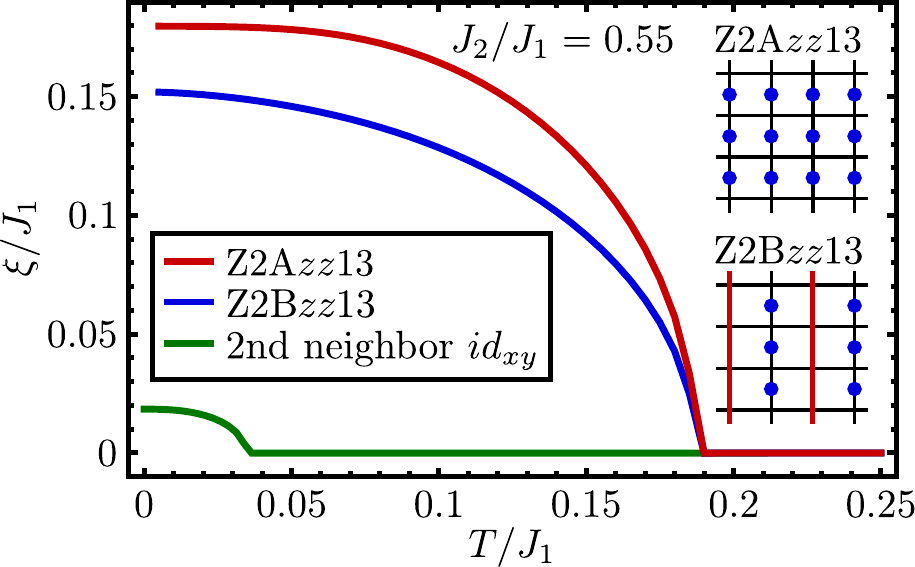}
\caption{Total amplitudes $\xi=(|t|^2+|\Delta|^2)^{1/2}$ from plain mean-field [Eq.~(\ref{bare_mf})] for the Z2A$zz$13 and Z2B$zz$13 states on the square lattice with $J_2/J_1=0.55$. Also shown is the amplitude of the imaginary second neighbor $id_{xy}$ pairing which appears on top of the Z2A$zz$13 state. Note that the plain mean-field amplitudes are defined as a function of temperature instead of $\Lambda$. (Inset) Illustration of both states where red lines (blue dots) denote negative hoppings (pairings) while otherwise the amplitudes are positive.\label{fig:bare_mf_square}}
\end{figure}

{\em $J_1$\textendash$J_2$ square lattice Heisenberg model}. We first present results for the $J_1$\textendash$J_2$ Heisenberg model on the 2D square lattice with antiferromagnetic $J_1$, $J_2>0$. In a regime around $J_2/J_1=0.55$, most numerical methods predict a non-magnetic ground state~\cite{Schulz-1992,Schulz-1996,Mambrini-2006,Jiang-2012,Gong-2014,Haghshenas-2018} representing a promising quantum spin liquid candidate. This has also been independently confirmed by PFFRG showing no indication of a magnetic instability during the RG flow of $\Gamma^\Lambda$ and only small valence-bond dimer susceptibilities~\cite{reuther10}.

To illustrate our combined PFFRG plus mean-field procedure and to benchmark with previous studies we first discuss the results of a plain mean-field treatment using Eq.~(\ref{bare_mf}) with bare couplings $J_{ij}$ and no renormalized self-energy $\Sigma^\Lambda$ inserted. We set $J_2/J_1=0.55$ and proceed successively by starting with nearest-neighbor amplitudes $t_1$, $\Delta_1$ and then add second neighbor terms $t_2$, $\Delta_2$. On the nearest-neighbor level, the two PSGs labelled Z2A$zz$13 and Z2B$zz$13 (see Ref.~\cite{wen02}) already cover all possible hopping/pairing terms (which can always be chosen real). The Z2A$zz$13 state has isotropic hopping $t_1$ and $d_{x^2-y^2}$ pairing $\Delta_1$ while Z2B$zz$13 exhibits staggered hopping $t_1$ and pairing $\Delta_1$ (see inset of Fig.~\ref{fig:bare_mf_square} for illustrations). Our plain mean-field results in Fig.~\ref{fig:bare_mf_square} indicate that below $T\approx0.19J_1$ the state Z2A$zz$13 acquires the largest amplitudes and, hence, appears to be preferred. Furthermore, we find equal hopping and pairing $t_1=\Delta_1$ for this state, in which case it can be shown that Z2A$zz$13 actually becomes gauge equivalent to the well-known SU(2) $\pi$-flux state (denoted SU2B$n$0 in Ref.~\cite{wen02}) with $\pi$-flux through every elementary square plaquette.
\begin{figure}[t]
\includegraphics[width=0.78\linewidth]{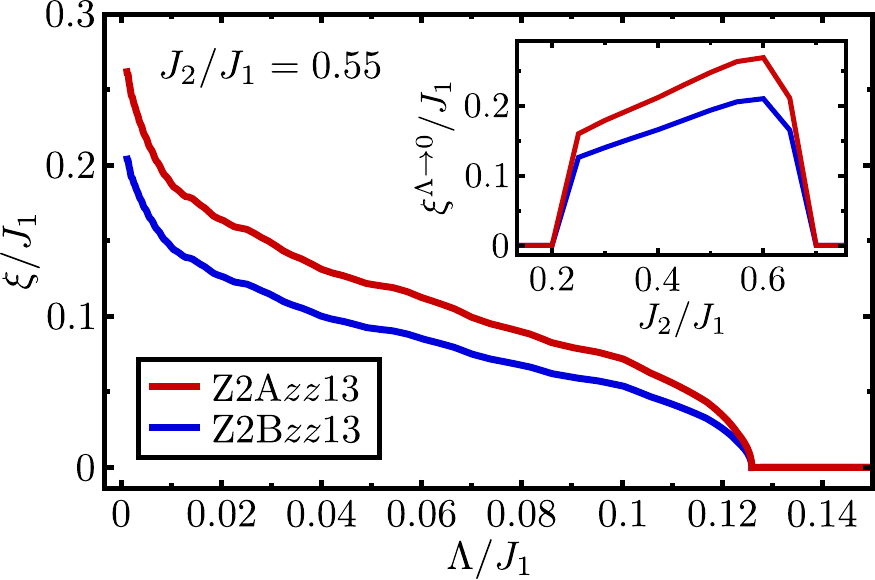}
\caption{Total amplitudes $\xi=(|t|^2+|\Delta|^2)^{1/2}$ from full PFFRG plus mean-field [Eq.~(\ref{full_mf})] for the Z2A$zz$13 and Z2B$zz$13 states on the square lattice with $J_2/J_1=0.55$. The inset shows $\xi$ for both states at small $\Lambda\rightarrow0$ within a wider range of $J_2/J_1$.\label{fig:full_mf_square}}
\end{figure}

Next, we consider second neighbor terms on top of the preferred $\pi$-flux state. The only symmetry-allowed amplitudes are real isotropic or real $d_{xy}$ hoppings $t_2$. Interestingly, neither of the two become finite during the RG flow. In fact, the only finite (albeit small) second neighbor term we find is an {\it imaginary} $id_{xy}$ pairing $\Delta_2$ which breaks time reversal symmetry, hence, signaling a {\it chiral} spin liquid, see green line in Fig.~\ref{fig:bare_mf_square}. Given the smallness of $\Delta_2$ we did not attempt to add further neighbor terms. Most importantly, this chiral state agrees with the one identified in the mean-field treatment of Ref.~\cite{wen02} confirming the correctness of our considerations.

\begin{figure}[t]
\includegraphics[width=0.99\linewidth]{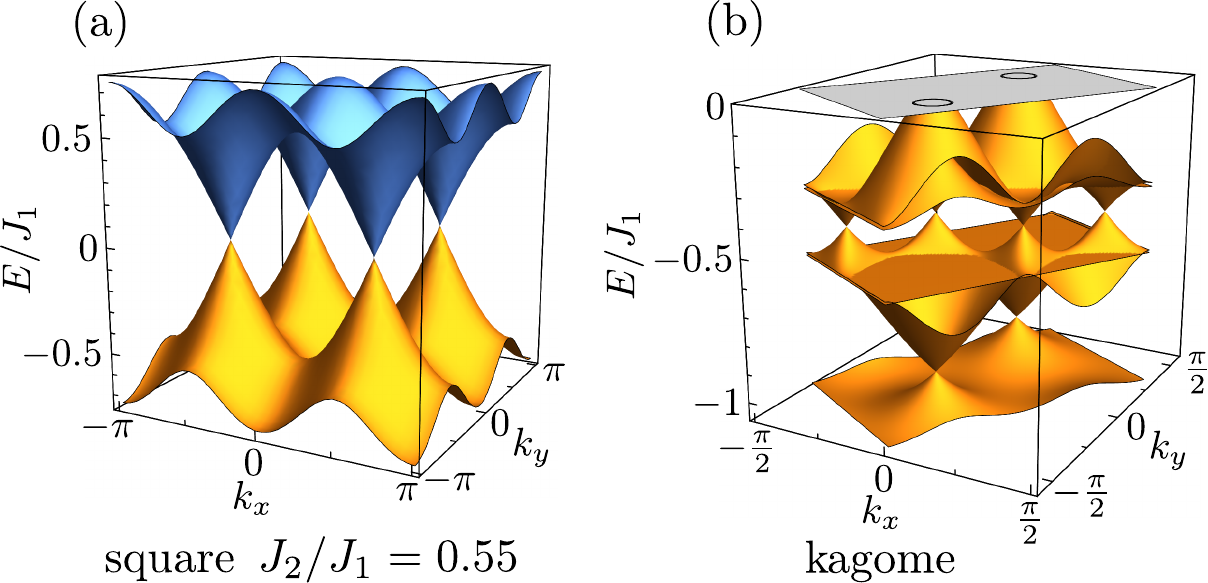}
\caption{Spinon bands from PFFRG plus Fock mean-field in the limit $\Lambda\rightarrow0$ for (a) the Z2A$zz$13 state of the square lattice Heisenberg antiferromagnet and (b) the $\mathds{Z}_2[0,\pi]\alpha$ state of the kagome Heisenberg antiferromagnet. In (b), only the negative part of the spectrum is shown while the positive part is the exact particle-hole transformed counterpart. The gray region is the first Brillouin zone with the Fermi surface indicated by black rings (the nearest-neighbor lattice constant is always set to one).\label{fig:spinon_bands}}
\end{figure}

We now repeat this analysis for the full PFFRG plus mean-field approach. Our results for the nearest-neighbor amplitudes of the Z2A$zz$13 and Z2B$zz$13 states are shown in Fig.~\ref{fig:full_mf_square}. While the shape of the curves is qualitatively different compared to our plain mean-field study, the overall result remains rather unchanged, particularly, we again find the Z2A$zz$13 state to be preferred. As before, the nearest-neighbor amplitudes are identical ($t_1=\Delta_1$) which suggests a SU(2) $\pi$-flux state. Considering longer-range terms, we do not detect any of the aforementioned second-neighbor amplitudes, not even the chiral $id_{xy}$ term. Hence, we conclude that the previous detection of a chiral spin liquid was an artefact of the plain mean-field treatment. As shown in Fig.~\ref{fig:spinon_bands}, the SU(2) $\pi$-flux state which we propose for this system features four gapless spinon Dirac cones with nodal Fermi points at ${\mathbf k}=(\pm\frac{\pi}{2},\pm\frac{\pi}{2})$. We also calculated the amplitude $\xi$ of both states within an extended region of $J_2/J_1$ and find that Z2A$zz$13 remains stable throughout the magnetically disordered regime (which ranges approximately from $J_2/J_1=0.4$ to $0.6$). Particularly, within the N\'eel (collinear) ordered phase at $J_2\lesssim 0.4 J_1$ ($J_2\gtrsim 0.6J_1$) the amplitudes $\xi$ are seen to decrease rapidly, confirming a direct connection to strong quantum fluctuations \footnote{Note that the vanishing of spinon hopping/pairing amplitudes does not yet signal a confinement-deconfinement transition. Rather spinon confinement requires the onset of long-range interactions between these quasiparticles.}.

It is instructive to benchmark these findings with VMC which compares variational energies of Gutzwiller-projected wave functions for different PSGs. In agreement with our analysis, VMC finds the Z2A$zz$13 state to be preferred as it yields the lowest variational energy~\cite{hu13}. However, differences occur on longer range $(\pm2,\pm2)$ bonds where VMC detects an additional finite $d_{xy}$ term that is absent in our analysis. 

{\em Kagome Heisenberg antiferromagnet}. The Heisenberg model on the kagome lattice with antiferromagnetic nearest-neighbor couplings $J_1$ is one of the most intriguing frustrated spin system. Early on it has been proposed to realize a quantum spin liquid ground state~\cite{Sachdev-1992,Lecheminant-1997}, however, its precise spinon properties (gapped versus gapless) and type of emergent gauge-field [U(1) versus $\mathds{Z}_2$] have remained a subject of ongoing debates~\cite{Yan-2011,Iqbal-2011,Depenbrock-2012,Jiang-2012b,Iqbal-2013,Iqbal-2014,Iqbal-2015b,Lauchli-2016,liao17,Mei-2017,He-2017,Iqbal-2018kag,Ralko-2018,Changlani-2018,zhu18}. Here, we apply the PFFRG plus Fock-mean-field approach to shed more light on these questions.

\begin{figure}[t]
\includegraphics[width=0.99\linewidth]{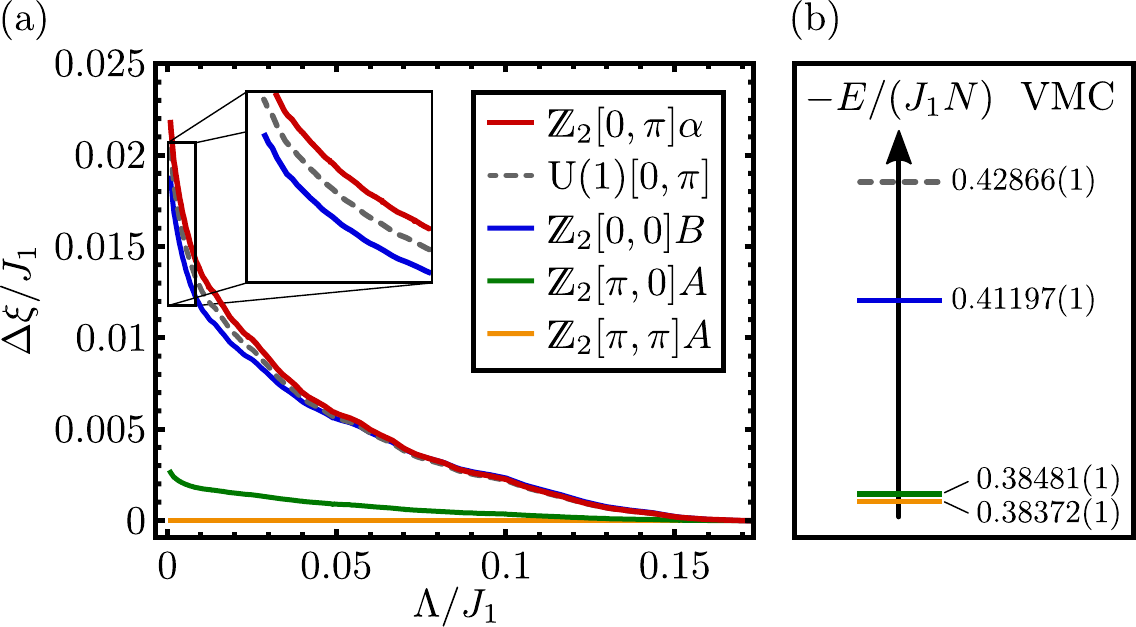}
\caption{(a) Nearest neighbor amplitudes $\xi$ from full PFFRG plus mean-field for the states $\mathds{Z}_2[0,\pi]\alpha$, $\mathds{Z}_2[0,0]B$, $\mathds{Z}_2[\pi,0]A$, and $\mathds{Z}_2[\pi,\pi]A$ on the kagome lattice. Only differences $\Delta\xi$ to the lowest state $\mathds{Z}_2[\pi,\pi]A$ are shown (which at $\Lambda\rightarrow0$ has an {\it absolute} amplitude $\xi=0.29J_1$). Also plotted is the amplitude for the $\mathds{Z}_2[0,\pi]\alpha$ ansatz when omitting pairing terms (effectively resulting in a U(1)$[0,\pi]$ state) see dashed line. (b) Corresponding variational energies per site from VMC~\cite{Ran-2007,Iqbal-2011a,Iqbal-2011}.\label{fig:full_mf_kagome}}
\end{figure}

According to a $\mathds{Z}_2$ PSG analysis~\cite{Lu-2011} there are four different ans\"atze covering all possible nearest-neighbor hopping and pairing terms, see Fig.~\ref{fig:psg_kagome}. In Ref.~\cite{Lu-2011}, they have been dubbed $\mathds{Z}_2[0,\pi]\alpha$, $\mathds{Z}_2[0,0]B$, $\mathds{Z}_2[\pi,0]A$, and $\mathds{Z}_2[\pi,\pi]A$, where the numbers in square brackets label the fluxes (of the parent U(1) state) from the hoppings through elementary triangles and hexagons, respectively~\footnote{For the $\mathds{Z}_2[0,\pi]\alpha$ and the $\mathds{Z}_2[0,0]B$ states, the first label in the square bracket denotes a ``0''-flux through both up- and down-triangles, while for the $\mathds{Z}_2[\pi,0]A$, and $\mathds{Z}_2[\pi,\pi]A$ states, the first label denotes a $\pi$-flux only through one-type of triangle, and a zero-flux through the other type. On the kagome lattice there are additional PSGs~\cite{Lu-2011}, such as those connected to (i) the parent U(1)$[0,\pi]$ state, namely, $\mathds{Z}_2[0,\pi]\beta$ (gapped), $\mathds{Z}_2[0,\pi]\gamma$ (gapless), and $\mathds{Z}_2[0,\pi]\delta$ (gapless), (ii) the parent U(1)$[0,0]$ state, namely, $\mathds{Z}_2[0,0]A$ (gapped), $\mathds{Z}_2[0,0]C$ (gapped), and $\mathds{Z}_2[0,0]D$ (gapped), etc, all of which realize a $\mathbb{Z}_{2}$ spin liquid via longer-range amplitudes. We do not find these amplitudes being generated such that these states reduce to one of the four states we have investigated.}. A first observation within PFFRG plus Fock mean-field is that the amplitudes $\xi$ for these ans\"atze are very similar in magnitude, such that in Fig.~\ref{fig:full_mf_kagome}(a) only the differences $\Delta\xi$ to the lowest state $\mathds{Z}_2[\pi,\pi]A$ are shown. This property is in line with the rather general observation of a dense manifold of competing low-energy states~\cite{Poilblanc-2011}. We identify the \emph{gapless} $\mathds{Z}_2[0,\pi]\alpha$ state as the preferred one where hoppings and pairing appear in a ratio $\frac{t_1}{\Delta_1}\approx2.48$ at small $\Lambda$, i.e., the pairing (which is responsible for the $\mathds{Z}_2$ gauge structure) is clearly subdominant. For the other three states pairing is not generated such that their gauge structure effectively remains U(1). We also considered onsite and up to third-neighbor terms on top of these ans\"atze, however, with amplitudes more than two orders of magnitude smaller, their effect can be safely neglected. Additionally, we investigated various chiral spin liquid ans\"atze~\cite{Ran-2007} which are found to reside in the gap between the $\mathds{Z}_2[\pi,0]A$ and $\mathds{Z}_2[0,0]B$ states in Fig.~\ref{fig:full_mf_kagome}(a)~\cite{supp}. We therefore propose the $\mathds{Z}_2[0,\pi]\alpha$ state with the gapless spinon band structure of Fig.~\ref{fig:spinon_bands}(b) to be realized as the ground state of the kagome Heisenberg antiferromagnet. The spinon dispersion shows approximate Dirac cones at small energies $|E|\lesssim 0.2J_1$, however, the Fermi level does not exactly intersect with the nodal points but rather cuts out two small circular Fermi surfaces. While recent VMC~\cite{Iqbal-2013}, DMRG~\cite{He-2017,zhu18}, and tensor network~\cite{liao17} approaches support the view of gapless Dirac spinons, small remaining Fermi rings have so far not been observed.

We have performed additional VMC calculations for these ans\"atze to benchmark our findings. Remarkably, the size of the amplitudes $\xi$ obtained from PFFRG and the variational energies $E$ from VMC~\cite{Ran-2007,Iqbal-2011a,Iqbal-2011} show the same sequence [Fig.~\ref{fig:full_mf_kagome}(b)] where even the size of the gaps between states appear similar. A clear difference, however, occurs for the favored $\mathds{Z}_2[0,\pi]\alpha$ state where VMC finds~\cite{Iqbal-2011} the lowest variational energy in the {\it absence} of pairing (which corresponds to the U(1)$[0,\pi]$ Dirac ansatz). We have also calculated the amplitude $\xi$ for this U(1)$[0,\pi]$ state and find it slightly below the $\mathds{Z}_2[0,\pi]\alpha$ ansatze [dashed line in Fig.~\ref{fig:full_mf_kagome}(a)]. Interestingly, the striking similarities in the sequence of amplitudes $\xi$ and variational energies $E$ also persists when including chiral states~\cite{supp}.

{\em Discussion and conclusion}. We have applied a new methodological combination of PFFRG and Fock-mean-field theory to calculate free spinon theories for quantum spin liquids in antiferromagnetic Heisenberg square and kagome lattice models. For the supposed quantum spin liquid on the square lattice we propose a SU(2) $\pi$-flux state with gapless Dirac spinons. For the kagome Heisenberg antiferromagnet our results indicate a gapless $\mathds{Z}_2$ spin liquid with two small circular Fermi surfaces. While we find an overall good agreement with VMC (particularly concerning the energetic hierarchy of states) certain differences are also revealed. Particularly, for the $\pi$-flux state on the square lattice, VMC finds~\cite{hu13} additional longer-range pairing which is absent in our results. Conversely, for the kagome spin liquid we identify additional nearest-neighbor pairing which VMC does not find~\cite{Iqbal-2011}.

It is important to emphasize that comparisons between PFFRG plus Fock mean-field and VMC generally need to be interpreted with caution since both methods calculate rather different quantities. While PFFRG plus Fock mean-field computes free spinon theories for quantum spin liquids, it does not determine ground-state energies. On the other hand, VMC calculates variational ground-state energies but the underlying free fermion models do not necessarily describe the actual spinon excitations of the system. Stated differently, agreement is {\it a priori} not expected and the Gutzwiller projected wave functions of our free spinon models might not even have good variational energies. The deeper reason for such conceptual differences is rooted in short range spinon-spinon interactions (mediated by gauge fluctuations) which are not determined within our approach. We would only expect good variational energies when they are computed from the ground state of an effective model {\it including} such interactions. Conversely, if good agreement between both methods is found (as for the models studied above) this might indicate weak spinon-spinon interactions possibly due to large flux (vison) gaps.

Finally, we note that the applicability of our method goes far beyond the systems studied here. Given the flexibility of the PFFRG, extensions towards three-dimensional systems as well as longer range and/or anisotropic frustrated spin interactions are feasible. Further possibilities are provided via extensions incorporating triplet pairing ans\"atze so as to describe spin and/or lattice-nematics, all promising a wealth of future applications.

{\em Acknowledgements}. The authors thank J\"{o}rg Behrmann, Sebastian Diehl, Max Geier, Achim Rosch, Ronny Thomale, and Simon Trebst for fruitful discussions. The numerics for the PFFRG were performed on the yoshi cluster at Freie Universit\"at Berlin. This work was partially supported by the German Research Foundation within the CRC 183 (project A02). J.R. acknowledges support by the Freie Universit\"at Berlin within the Excellence Initiative of the German Research Foundation. Y.I. acknowledges the kind hospitality of the Helmholtz-Zentrum f\"{u}r Materialien und Energie, Berlin, where part of the work was carried out. 


%

\newpage



\newcommand{\beginsupplement}{%
        \setcounter{table}{0}
        \renewcommand{\thetable}{S\arabic{table}}%
        \setcounter{figure}{0}
        \renewcommand{\thefigure}{S\arabic{figure}}%
        \setcounter{equation}{0}
        \renewcommand{\theequation}{S\arabic{equation}}%
        \setcounter{page}{1}
     }
 
 \bibliographystyle{apsrev4-1}
\renewcommand*{\citenumfont}[1]{#1}
\renewcommand*{\bibnumfmt}[1]{[#1]}
 
\newcommand\blankpage{%
    \null
    \thispagestyle{empty}%
    \addtocounter{page}{-1}%
    \newpage}


\chead{{\large \bf{--- Supplemental Material ---\\Characterization of quantum spin liquids and their spinon band structures via functional renormalization}}}

\thispagestyle{fancy}

\beginsupplement

{\em Methodological details}. Here we provide more details about the exact form of the Fock-mean-field equations and explain how to solve them. The full expression for the matrix Green's function in real space and imaginary time is
\begin{align}
 G_{ij}(\tau,\tau')&=-\left \langle T_{\tau} \hat{\psi}_i(\tau) \hat{\psi}^{\dagger}_j(\tau') \right \rangle \nonumber \\ &= \left (\begin{array}{c c} \mathcal{G}_{ij}(\tau,\tau') & \mathcal{F}_{ij}(\tau,\tau') \\ \mathcal{F}^{\ast}_{ij}(\tau',\tau)  & -\mathcal{G}_{ji}(\tau',\tau) \end{array} \right ), \\ \mathcal{G}_{ij}(\tau,\tau')&=-\left \langle T_{\tau} \hat{f}_{i\uparrow}(\tau) \hat{f}^{\dagger}_{j \uparrow}(\tau') \right \rangle, \\ \mathcal{F}_{ij}(\tau,\tau')&=-\left \langle T_{\tau} \hat{f}_{i \uparrow}(\tau) \hat{f}_{j \downarrow}(\tau') \right \rangle.
\end{align}
The self-consistent mean-field equation for the bare interaction in Matsubara and momentum space reads
\begin{align}\label{eq:FinalBare}
 G_{\mathbf k}(\omega)&= \left ( i\omega \mathds{1}_{2\times 2} -u_{\mathbf k} \right)^{-1},
  \\ u_{\mathbf k}&= - \lim\limits_{\lambda \rightarrow 0} \frac{3}{4 \beta}\sum\limits_{\mathbf q}  \sum\limits_{\omega} J_{{\mathbf k}-{\mathbf q}} G^{\lambda}_{\mathbf q}(\omega), \\  G^{\lambda}_{\mathbf q}(\omega)&=e^{i\omega \lambda \sigma^z}  G_{\mathbf q}(\omega),
\end{align}
where $\sum\limits_{\mathbf q}$ sums over the Brillouin zone and $\sigma^z$ is the third Pauli matrix acting in Nambu space.

When using renormalized vertices, we evaluate the Green's function via
\begin{equation}\label{eq:FinalFRG}
 G^{\Lambda}_{\mathbf k}(\omega)= \theta(|\omega|-\Lambda) \left[\left (i \omega-\Sigma^{\Lambda}(\omega) \right) \mathds{1}_{2\times 2} -u^{\Lambda}_{\mathbf k}\right]^{-1},
\end{equation}
where the hopping and pairing terms are determined from
\begin{equation}
 u^{\Lambda}_{\mathbf k}= -\int\limits_{-\infty}^{\infty}\frac{d\omega}{2\pi}  \sum\limits_{\mathbf q} \left(3\Gamma^{\Lambda \,s}_{{\mathbf k}-{\mathbf q}}(s,t,u) +\Gamma^{\Lambda \,d}_{{\mathbf k}-{\mathbf q}}(s,t,u)\right) G^{\Lambda}_{\mathbf q}(\omega).
\end{equation}
Here, we have used the vertex parameterization into spin and density channels $\Gamma^{\Lambda \,s}$ and $\Gamma^{\Lambda \,d}$ from Ref.~\cite{reuther10}. Furthermore, we performed the $T\rightarrow 0$ limit and assumed $u^{\Lambda}_{\mathbf k}$ to be constant in Matsubara frequency, \textit{i.e.}, all hoppings and pairings are instantaneous in imaginary time $\left[\propto \delta(\tau-\tau')\right]$.

For non-Bravais lattices and for those ans\"atze that do not obey the translation invariance of their underlying lattice (which is the case for most of the ans\"atze studied here), we have to extend the matrix structure of the self-consistent equations' constituents into sublattice space. Labeling the sites within a unit cell of the mean-field ansatz $a,b,\,\dots\,$, we consider the quantities

\begin{align}
 J_{ij}&= \left (\begin{array}{c c c} J^{aa}_{ij} & J^{ab}_{ij} & \cdots \\ J^{ba}_{ij} & J^{bb}_{ij} & \cdots \\ \vdots &\vdots &\ddots \end{array} \right ),  \\ u_{ij}&= \left (\begin{array}{c c c c c c} t^{aa \, \ast}_{ij} & t^{ab \, \ast}_{ij} & \cdots & \Delta^{aa}_{ij} & \Delta^{ab}_{ij} & \cdots \\ t^{ba \, \ast}_{ij} & t^{bb \, \ast}_{ij} & \cdots & \Delta^{ba}_{ij} & \Delta^{bb}_{ij} & \cdots \\ \vdots &\vdots &\ddots & \vdots &\vdots &\ddots \\ \Delta^{aa \, \ast}_{ij} & \Delta^{ab \, \ast}_{ij} & \cdots  & -t^{aa}_{ij} & -t^{ab}_{ij} & \cdots \\ \Delta^{ba \, \ast}_{ij} & \Delta^{bb \, \ast}_{ij} & \cdots  & -t^{ba}_{ij} & -t^{bb}_{ij} & \cdots \\ \vdots &\vdots &\ddots & \vdots &\vdots &\ddots \end{array} \right ), \\ G_{ij}&= \left (\begin{array}{c c c c c c} \mathcal{G}^{aa}_{ij} & \mathcal{G}^{ab}_{ij} & \cdots & \mathcal{F}^{aa}_{ij} & \mathcal{F}^{ab}_{ij} & \cdots \\ \mathcal{G}^{ba}_{ij} & \mathcal{G}^{bb}_{ij} & \cdots & \mathcal{F}^{ba}_{ij} & \mathcal{F}^{bb}_{ij} & \cdots \\ \vdots &\vdots &\ddots & \vdots &\vdots &\ddots \\ \mathcal{F}^{aa \, \ast}_{ij} & \mathcal{F}^{ab \, \ast}_{ij} & \cdots  & -\mathcal{G}^{aa}_{ji} & -\mathcal{G}^{ba}_{ji} & \cdots \\ \mathcal{F}^{ba \, \ast}_{ij} & \mathcal{F}^{bb \, \ast}_{ij} & \cdots  & -\mathcal{G}^{ab}_{ji} & -\mathcal{G}^{bb}_{ji} & \cdots \\ \vdots &\vdots &\ddots & \vdots &\vdots &\ddots \end{array} \right ).
\end{align}
$\Gamma^{\Lambda}_{ij}$ is represented the same way as $J_{ij}$ and both matrices are proportional to the unit matrix in Nambu space. The products $J_{{\mathbf k}-{\mathbf q}}G_{\mathbf q}$ and $\Gamma^{\Lambda}_{{\mathbf k}-{\mathbf q}}G_{\mathbf q}$ have to be performed element wise (no matrix multiplication) in sublattice space. However, the sublattice space is relevant for the matrix inversion in the computation of $G_{\mathbf q}$. \par
{\em Numerical implementation}. The PFFRG calculations for the renormalized vertices were performed as described in Refs.~\cite{reuther10,Suttner-2014}. We used a mesh of $100$ Matsubara frequencies and limited the range of the vertex functions to clusters of $441$ ($125$) lattice sites for the square (kagome) lattice.
\begin{figure}[t]
\includegraphics[width=0.9\linewidth]{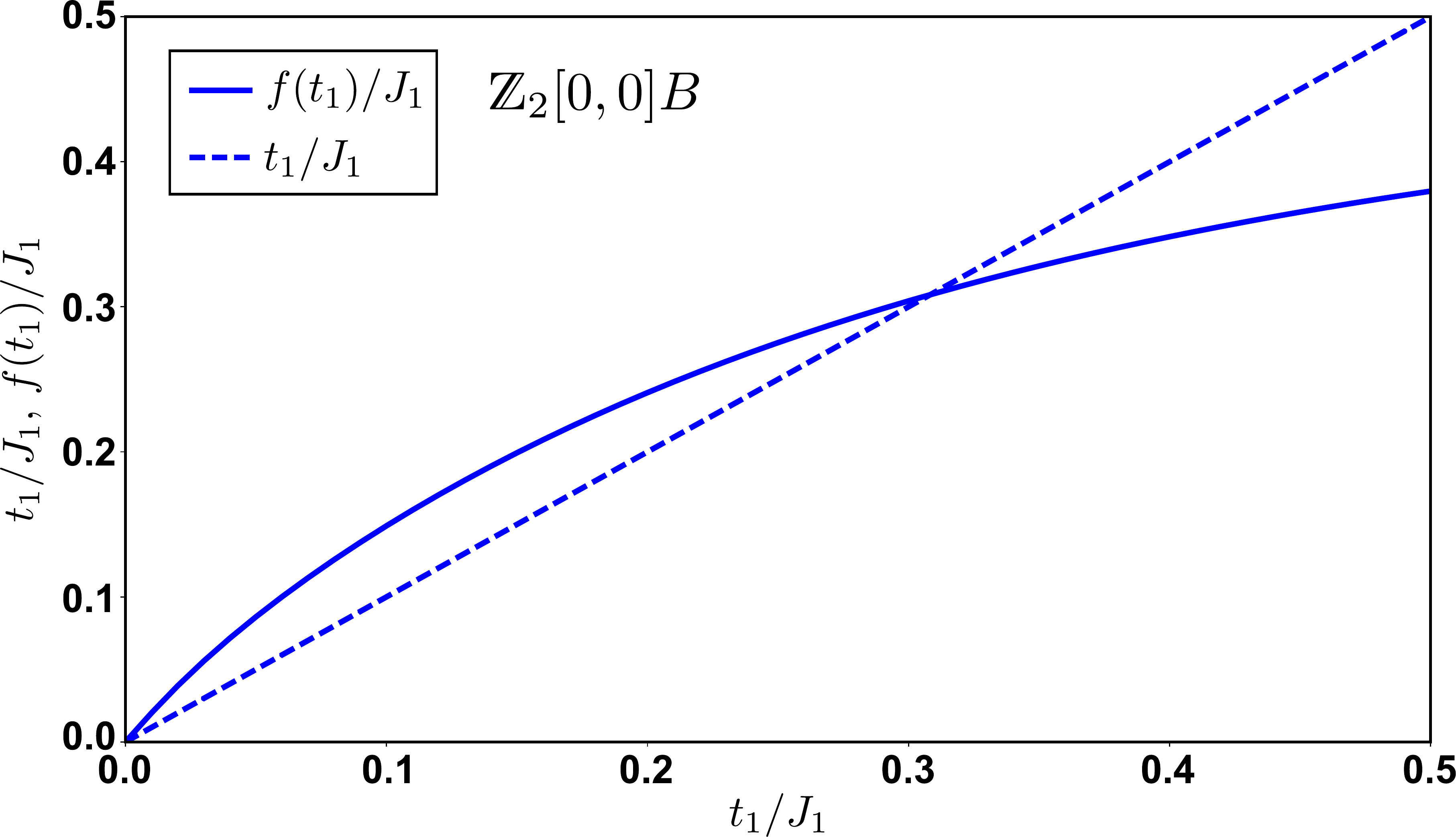}
\caption{Self-consistent equation with full PFFRG vertices for the $\mathds{Z}_2[0,0]B$ ansatz on the kagome lattice in the limit $\Lambda\rightarrow0$. Plotted are the left side (dotted line) and right side (full line) of the self-consistent equation (\ref{full_mf}) as a function of the nearest-neighbor hopping $t_1$. The right side of the flow equation is denoted $f(t_1)$.\label{fig:example}}
\end{figure}

For the evaluation of the self-consistent equations, the Matsubara sum can be performed analytically in the case where the frequency-independent bare interaction is used (mean-field calculation). The resulting terms are temperature dependent and the remaining momentum integrals were numerically evaluated using \textit{Wolfram Mathematica}. For the renormalized vertices ($T\rightarrow 0$ limit), the Matsubara integrals were performed on a similar frequency grid as for the PFFRG using $120$ discretized frequencies for each value of $\Lambda$. Furthermore, the momentum integrals were solved by assuming a grid with $225$ points for the full Brillouin zone of the square lattice and with $170$ points for the reduced Brillouin zone of the kagome lattice (corresponding to the six-site unit cell of the $\mathds{Z}_2[0,\pi]\alpha$ state, Fig.~\ref{fig:psg_kagome}). We have confirmed that further increasing the Matsubara frequency and $\mathbf k$ space resolutions does not affect the resulting amplitudes by more than $0.1\%$. As an example, we illustrate in Fig.~\ref{fig:example} the solution for the $\mathds{Z}_2[0,0]B$ ansatz on the kagome lattice where both sides of the self-consistent equation are plotted as a function of the nearest-neighbor hopping $t_1$. The intersection of both curves determines the solution.
\begin{figure}[t]
\includegraphics[width=0.99\linewidth]{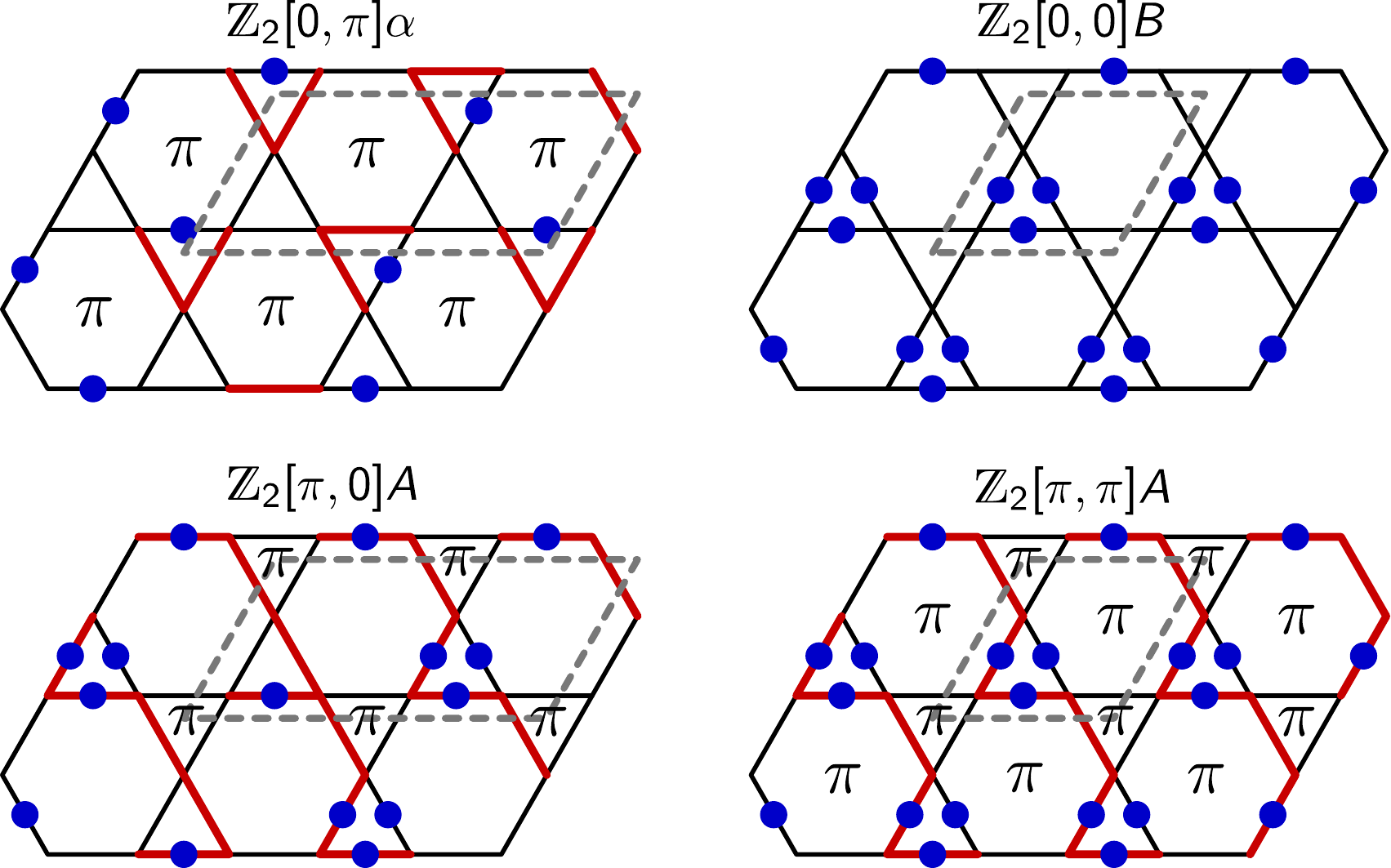}
\caption{Non-chiral nearest-neighbor ans\"atze on the kagome lattice. Each nearest-neighbor bond carries real hopping and real pairing amplitudes where the absolute value of the hoppings $|t_1|$ is the same on all bonds (the same also holds for the absolute values of the pairings $|\Delta_1|$). Red lines (blue dots) denote negative hoppings (negative pairings) while otherwise the amplitudes are positive. Labels $\pi$ indicate fluxes from the hoppings through respective plaquettes. Gray dashed lines illustrate the unit cell of the ansatz.\label{fig:psg_kagome}}
\end{figure}
\begin{figure}[t]
\includegraphics[width=0.99\linewidth]{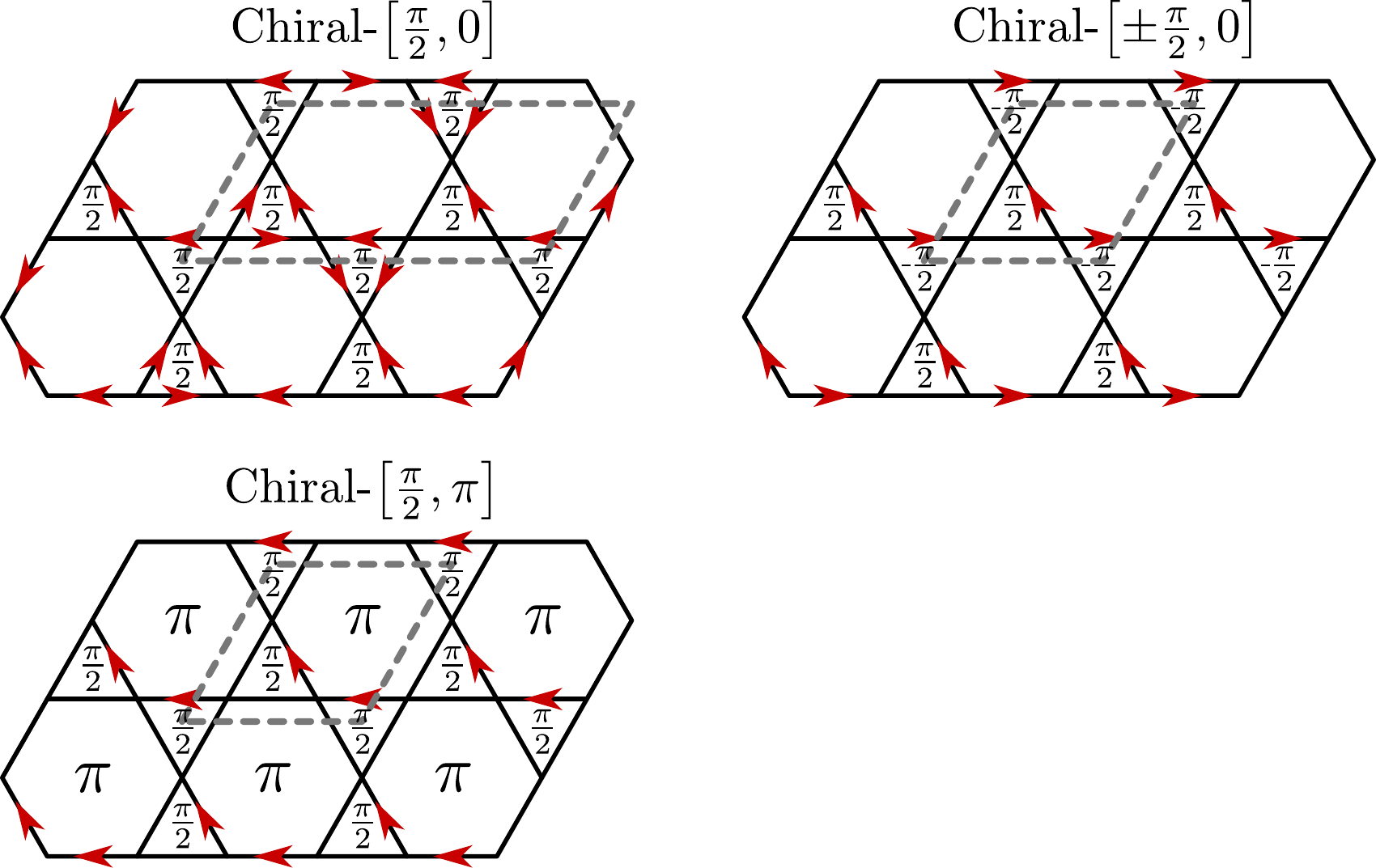}
\caption{Chiral nearest-neighbor ans\"atze on the kagome lattice. Each nearest-neighbor bond features hopping but no pairing amplitudes. Black bonds denote real and positive hopping $t_1>0$ while bonds with an arrow carry {\it imaginary} hopping $i t_1 f_i f_{j}^\dagger+\text{h.c.}$ (again with real $t_1>0$) where the arrows point from site $j$ to site $i$. Labels $\pm\frac{\pi}{2}$ and $\pi$ indicate fluxes through respective plaquettes. Gray dashed lines illustrate the unit cell of the ansatz.\label{fig:psg_kagome_chiral}}
\end{figure}

{\em Nearest neighbor PSG ans\"atze on the kagome lattice}.
In Fig.~\ref{fig:psg_kagome} we illustrate the four non-chiral nearest-neighbor PSG ans\"atze on the kagome lattice ($\mathds{Z}_2[0,\pi]\alpha$, $\mathds{Z}_2[0,0]B$, $\mathds{Z}_2[\pi,0]A$, and $\mathds{Z}_2[\pi,\pi]A$), all containing hopping and pairing amplitudes. These states are characterized by different flux patterns from the hoppings through the elementary triangles (first number in square brackets) and hexagons (second number in square brackets), as also indicated in the figure. Two of these ans\"atze ($\mathds{Z}_2[0,\pi]\alpha$ and $\mathds{Z}_2[\pi,0]A$) have an enlarged unit cell containing six sites instead of the three-site unit cell of the kagome lattice. Most importantly, however, due to the projective implementation of symmetries within a PSG treatment, the spin state itself still respects all lattice symmetries. On the nearest-neighbor level, the states $\mathds{Z}_2[\pi,0]A$ and $\mathds{Z}_2[\pi,\pi]A$ actually realize U(1) spin liquids, i.e., longer range pairings are required to break the gauge structure down to $\mathds{Z}_2$. Also note that among these four states, $\mathds{Z}_2[0,0]B$ is the only one with a finite spinon gap.
\begin{figure}
\includegraphics[width=0.99\linewidth]{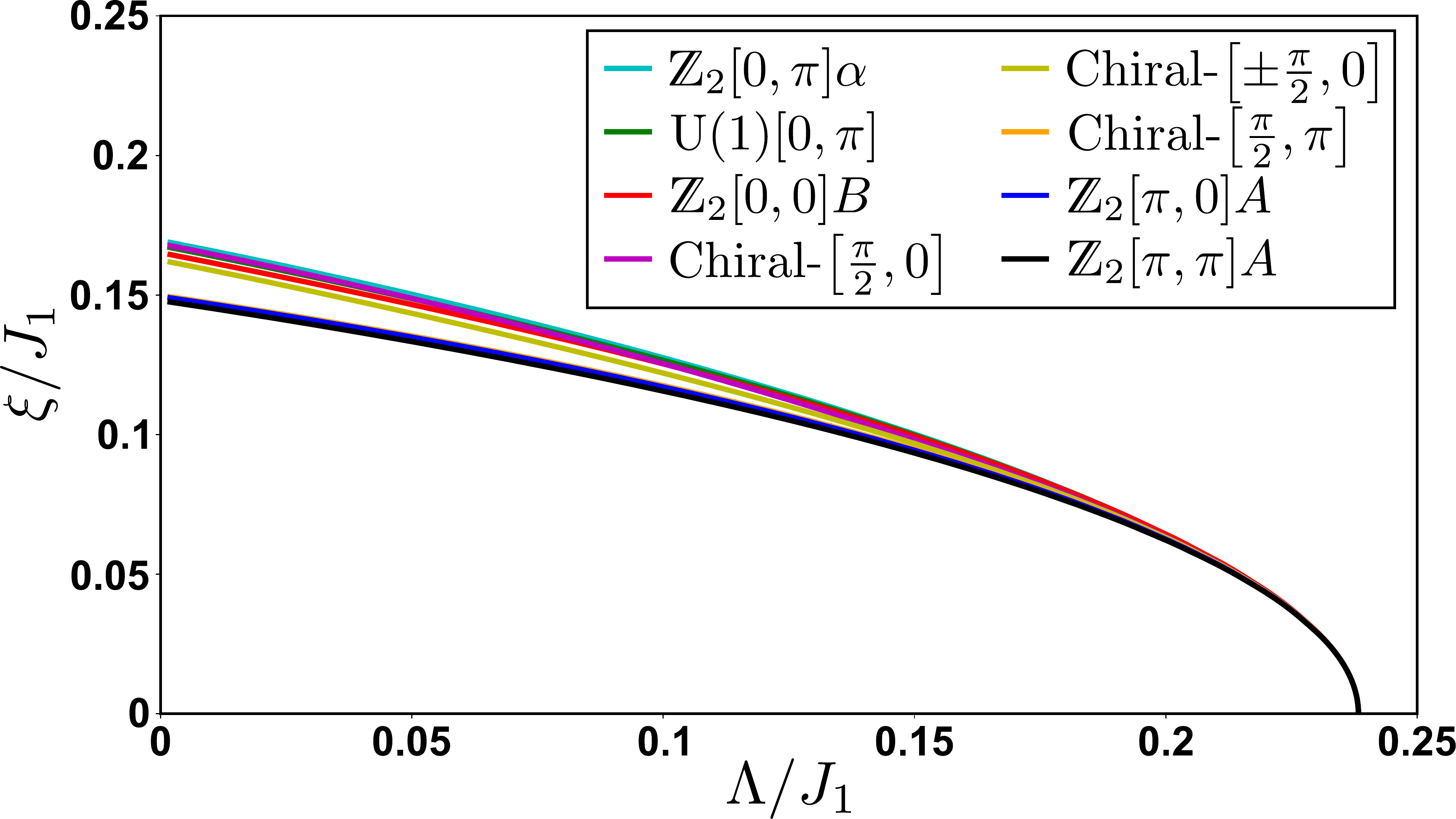}
\caption{Amplitudes $\xi$ from modified bare Fock-mean-field theory for chiral and non-chiral nearest-neighbor ans\"atze on the kagome lattice.\label{fig:bare_absolute}}
\end{figure}
\begin{figure}
\includegraphics[width=0.99\linewidth]{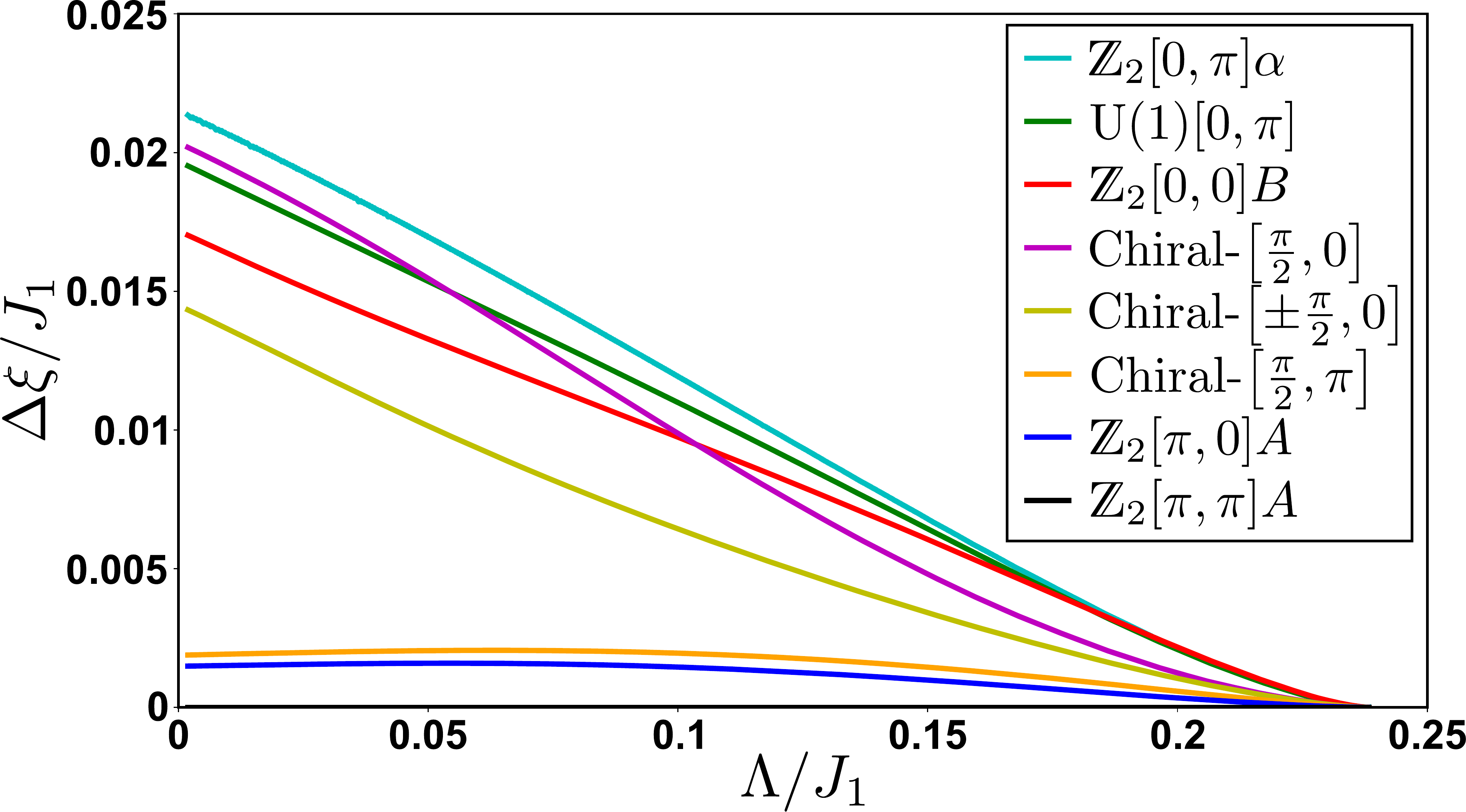}
\caption{Amplitudes $\Delta\xi$ from modified bare Fock-mean-field theory for chiral and non-chiral nearest-neighbor ans\"atze on the kagome lattice. The results are the same as in Fig.~\ref{fig:bare_absolute} but with amplitudes plotted relative to the lowest $\mathds{Z}_2[\pi,\pi]A$ state.\label{fig:bare_relative}}
\end{figure}
\begin{figure}
\includegraphics[width=0.99\linewidth]{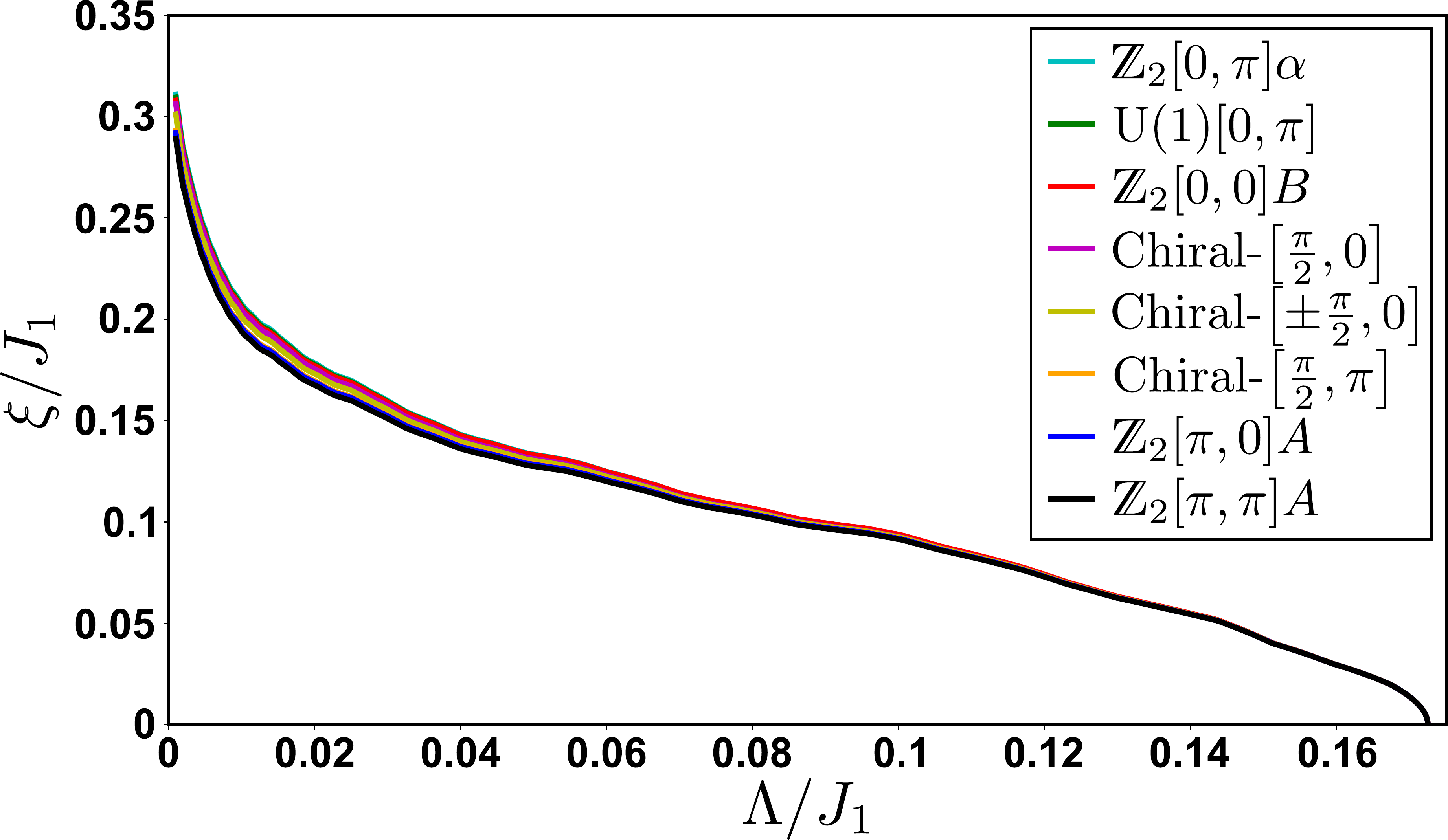}
\caption{Amplitudes $\xi$ from full PFFRG plus Fock-mean-field theory for chiral and non-chiral nearest-neighbor ans\"atze on the kagome lattice.\label{fig:full_absolute}}
\end{figure}

When relaxing time reversal invariance one obtains a plethora of chiral spin liquids~\cite{Bieri-2016}, and we investigate three promising candidate states , namely, the Chiral-$\left[\frac{\pi}{2},0\right]$, Chiral-$\left[\pm\frac{\pi}{2},0\right]$, and Chiral-$\left[\frac{\pi}{2},\pi\right]$ ans\"atze, as illustrated in Fig.~\ref{fig:psg_kagome_chiral}. Again, these states exhibit different flux patterns through triangles/hexagons where now staggered fluxes of $+\frac{\pi}{2}$ ($-\frac{\pi}{2}$) through up (down) triangles are also allowed. Note that the Chiral-$\left[\frac{\pi}{2},0\right]$ ansatz has a doubled unit cell, while the other two share the three-site geometrical unit cell of the kagome lattice.

{\em Mean-field solutions of chiral and non-chiral states on the kagome lattice}.
In addition to the results in the main part of the paper, here we show more details of the mean-field solutions on the kagome lattice, also including the chiral ans\"atze of Fig.~\ref{fig:psg_kagome_chiral}. In Figs.~\ref{fig:bare_absolute} and~\ref{fig:bare_relative}, we first plot the solutions from a bare mean-field theory. For better solvability of the mean-field equations we use a modified scheme where instead of the temperature $T$ as in Fig.~\ref{fig:bare_mf_square} we consider $T\rightarrow 0$ and use the $\Lambda$-regularized propagator $G_0^\Lambda$ together with $\Sigma^{\Lambda}(\omega)=0$ and the bare interactions $J_{ij}$ in the mean-field approach. Fig.~\ref{fig:bare_absolute} shows the absolute size of the amplitudes $\xi$ and Fig.~\ref{fig:bare_relative} shows -- for better visibility -- the amplitudes $\Delta\xi$ relative to the lowest $\mathds{Z}_2[\pi,\pi]A$ ansatz. Plotted are the quantity $\xi$ (or $\Delta\xi$) for the four non-chiral and three chiral nearest-neighbor ans\"atze as well as the solution for the hopping amplitude of the U(1)$[0,\pi]$ state which is obtained when omitting the pairing of the $\mathds{Z}_2[0,\pi]\alpha$ ansatz. As can be seen, the $\mathds{Z}_2[0,\pi]\alpha$ state appears preferred while the chiral ans\"atze have amplitudes with intermediate strengths. In particular, it is worth noting that the Chiral-$[\frac{\pi}{2},0]$ ansatz has a higher amplitude on the mean-field level compared to the $U(1)[0,\pi]$-state in agreement with the findings in Refs.~\cite{Zeng-1991,Hastings-2000} wherein the Chiral-$[\frac{\pi}{2},0]$ state was found to have a lower mean-field energy compared to the $U(1)[0,\pi]$-state.

\begin{figure}[b]
\includegraphics[width=0.99\linewidth]{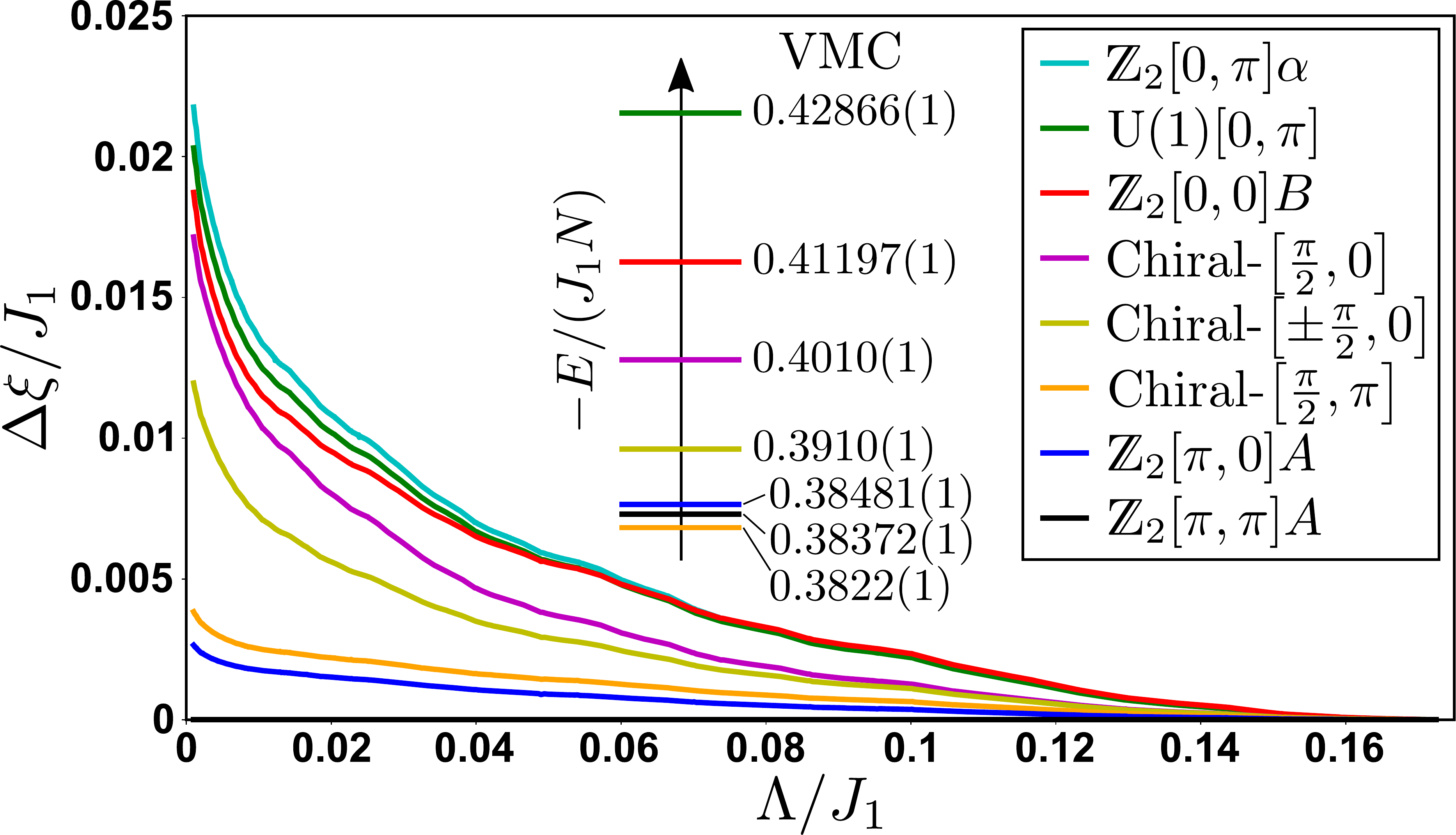}
\caption{(Main panel) Amplitudes $\Delta\xi$ from full PFFRG plus Fock-mean-field theory for chiral and non-chiral nearest-neighbor ans\"atze on the kagome lattice. Same as Fig.~\ref{fig:full_absolute} but with amplitudes plotted relative to the lowest $\mathds{Z}_2[\pi,\pi]A$ state. (Inset) Corresponding variational energies per site from VMC~\cite{Ran-2007,Iqbal-2011a,Iqbal-2011}.\label{fig:full_relative}}
\end{figure}

For comparison we also show the corresponding results for a full PFFRG plus mean-field treatment, again for absolute amplitudes (Fig.~\ref{fig:full_absolute}) and for relative amplitudes (Fig.~\ref{fig:full_relative}). Note that Fig.~\ref{fig:full_relative} contains the same results as Fig.~\ref{fig:full_mf_kagome} of the main text but with amplitudes for the chiral states added. We observe that compared to the plain mean-field results of Fig.~\ref{fig:bare_absolute} the curves now lie somewhat closer together indicating that quantum fluctuations lead to a stronger competition between different states. Overall, we observe larger amplitudes in Fig.~\ref{fig:full_absolute} compared to Fig.~\ref{fig:bare_absolute}, due to a  pronounced increase at small $\Lambda$. The sequence of amplitudes in Figs.~\ref{fig:bare_relative} and~\ref{fig:full_relative} has remained rather unchanged when including full PFFRG vertices except for the U(1)$[0,\pi]$, $\mathds{Z}_2[0,0]B$, and Chiral-$\left[\frac{\pi}{2},0\right]$ states which have swapped their positions, similar to what is observed in VMC studies~\cite{Ran-2007}. Particularly, the crossing of amplitudes in Fig.~\ref{fig:bare_relative} caused by the Chiral-$\left[\frac{\pi}{2},0\right]$ ansatz has disappeared in Fig.~\ref{fig:full_relative}. In total, the chiral states in Fig.~\ref{fig:full_relative} are located in the gap of amplitudes between the $\mathds{Z}_2[\pi,0]A$ and $\mathds{Z}_2[0,0]B$ states.

We finally compare the sequence of states in Fig.~\ref{fig:full_relative} with the variational energies from VMC, where in addition to Fig.~\ref{fig:full_mf_kagome} of the main text, we also include results for the chiral states (the variational energies for the chiral states are taken from Ref.~\cite{Ran-2007}). Interestingly, we again find striking similarities in the hierarchy of our amplitudes and the energies from VMC. Indeed, the only obvious difference is that in our results the Chiral-$\left[\frac{\pi}{2},\pi\right]$ state has the third smallest amplitude among all eight tested states while it has the highest variational energy. However, the energy scales associated with these differences are comparatively small.
\end{document}